\documentclass[aps,prl,onecolumn,superscriptaddress]{revtex4}

\usepackage[dvips]{graphicx}
\newcommand{\be}{\begin{displaymath}}
\newcommand{\bn}{\begin{equation}}

\newcommand{\en}{\end{equation}}
\newcommand{\ee}{\end{displaymath}}

\usepackage{times}
\usepackage{amssymb}
\usepackage{amsmath}
\usepackage{url,hyperref}

\begin{document}
\title{Resilience of quasi-isodynamic stellarators against trapped-particle
instabilities}

\author{J.~H.~E.~Proll} \author{P.~Helander}
\affiliation{Max-Planck-Institut f\"ur Plasmaphysik, EURATOM Association,
Teilinstitut Greifswald, Wendelsteinstra{\ss}e 1, 17491 Greifswald, Germany} 

\author{J.~W.~Connor}
\affiliation{Culham Centre for Fusion Energy, Abingdon OX14 3DB, United Kingdom}
\affiliation{Imperial College of Science, Technology and Medicine, London  SW7 2BZ,
United Kingdom}

\author{G.~G.~Plunk}
\affiliation{Max-Planck-Institut f\"ur Plasmaphysik, EURATOM Association,
Teilinstitut Greifswald, Wendelsteinstra{\ss}e 1, 17491 Greifswald, Germany}

\date{May 8, 2012}

\begin{abstract}
It is shown that in perfectly quasi-isodynamic stellarators, trapped particles with a bounce
frequency much higher than the frequency of the instability are stabilizing
in the electrostatic and collisionless limit. The collisionless trapped-particle
instability is therefore stable as well as the ordinary electron-density-gradient-driven
trapped-electron mode. This result follows from the energy balance of electrostatic
instabilities and is thus independent of all other details of the magnetic geometry.

\end{abstract}

\maketitle
\normalsize
Stellarators seek to confine fusion plasmas by means of a three-dimensionally shaped
magnetic field. In recent years, the art of optimizing this field to improve plasma
performance has taken great strides. In particular, it has proven possible to shape
the magnetic field in such a way that the collisional (so-called neoclassical)
transport is reduced almost to the level of axisymmetric devices. An important
question that then arises is how this optimization affects the properties of
microinstabilities and the turbulence they tend to cause. In tokamaks, most of this
turbulence is driven by ion and electron-temperature gradient (ITG and ETG) modes
and by the trapped-electron mode (TEM). In this Letter, we demonstrate that one of
the most important classes of orbit-optimized stellarators, so-called
quasi-isodynamic ones, is automatically immune to the ordinary TEM and to all lower-frequency electrostatic instabilities, if the temperature
gradients are small enough compared with the density gradient and collisions can be ignored. Quasi-isodynamic
stellarators could therefore benefit from reduced transport both in the neoclassical
and turbulent channels. 

A toroidal magnetic field $\bf B$ is quasi-isodynamic when the contours of constant
$B = |{\textbf B}|$ are poloidally (but not toroidally) closed and all collisionless orbits are perfectly
confined \cite{HN, Nuhrenberg2010}. Thus, if ${\bf B} = \nabla \psi \times \nabla
\alpha$, where $\psi$ denotes the toroidal flux, then the radial drift should vanish
when averaged over the bounce time $\tau_b$,
        $$ \frac{1}{\tau_b} \int_0^{\tau_b} {\bf v}_d \cdot \nabla \psi \; dt = 0. $$
The parallel adiabatic invariant,
        $$ J = \int_{l_1}^{l_2} mv_\| dl, $$
where the integral is taken along the field between two successive bounce points, is
then constant on flux surfaces, i.e., $J$ depends on $\psi$, the energy and magnetic
moment of the particle but is independent of $\alpha$. Wendelstein 7-X is the first
stellarator to approach quasi-isodynamicity, and substantially more quasi-isodynamic
configurations have been found computationally in the last few years
\cite{Nuhrenberg2010, Subbotin2006, Mikhailov2009}. 

These devices are so-called maximum-$J$ configurations, where $J$ has a maximum on
the magnetic axis and $\partial J / \partial \psi < 0$. In 1968, Rosenbluth
\cite{Rosenbluth1968} had already noticed that the maximum-$J$ property is beneficial for the
stability of interchange modes with frequencies above the drift frequency but below the bounce frequency of all
plasma constituents (see also Refs. \cite{Rutherford1968a, Antonsen1987}). For typical microinstabilities (except the ETG mode), this condition holds for the electrons but not for the ions. Nevertheless, considering the full gyrokinetic system of equations, we show in this Letter that stability prevails far beyond the limit considered by Rosenbluth. 

The physical reason for this remarkable stability has to do with the direction of
the precessional drift of the trapped particles. If the wave vector perpendicular to
the magnetic field is $\textbf{k}_\bot = k_{\alpha} \nabla \alpha + k_{\psi}\nabla
\psi$, we define the magnetic drift frequency $\omega_{da} = \textbf{k}_{\bot}\cdot
\textbf{v}_{da}$ and the drift wave frequency $\omega_{*a} = (T_a k_{\alpha}/e_a) d
\ln n_a / d \psi$ for each particle species $a$ in the usual way, where the density
$n_a$ and temperature $T_a$ are constant on flux-surfaces, and $\textbf{v}_{da} =
\hat{\textbf{b}} \times ( (v_{\bot}^2/2)\nabla \ln B +
v_{\|}^2\boldsymbol{\kappa})/\Omega_a$ denotes the drift velocity, with $\hat{\textbf{b}} =
{\bf B}/|{\bf B}|$ the unit tangent vector and $\boldsymbol{\kappa} =
\hat{\textbf{b}} \cdot \nabla \hat{\textbf{b}}$ the curvature vector of the magnetic
field. The precession frequency then becomes
$$ \overline{\omega_{da}}= k_{\alpha}\overline{\nabla \alpha \cdot \textbf{v}_{da}}  + 
k_{\psi}\overline{\nabla \psi \cdot \textbf{v}_{da}},$$
where an overbar denotes the bounce average. By design the last term vanishes for
quasi-isodynamic configurations. The remaining term can be expressed as a derivative
of the parallel adiabatic invariant, taken at fixed energy and magnetic moment, 
$$ k_{\alpha}\overline{\nabla \alpha \cdot \textbf{v}_{da}} = -
\frac{k_{\alpha}}{Ze_a \tau_{b a}} \frac{\partial J}{\partial \psi}, $$
and the product of the precession and drift wave frequencies \cite{Rosenbluth1971b},
$$
\omega_{*a}\cdot \overline{\omega_{da}}=- k_{\alpha}^2 \dfrac{T_a }{Z e_a^2  \tau_{b
a}}\dfrac{d \ln n_a}{d \psi} \frac{\partial J}{\partial \psi},
$$
is therefore negative in a maximum-$J$-configuration with a density that
increases toward the plasma center,
\begin{equation}
\omega_{*a}\cdot \overline{\omega_{da}}< 0.
\label{eq:qi}
\end{equation}
Typically, trapped-particle instabilities rely on the resonance between these two frequencies, which occurs due to so-called ``bad'' curvature. In quasi-isodynamic configurations, however, we see that trapped particles have average ``good'' curvature and thus exert a stabilizing influence.
To demonstrate this mathematically, we proceed from the gyrokinetic system of equations in ballooning space,
\begin{equation}
i v_{\|}\nabla_{\|}g_a+(\omega -\omega_{da})g_a = \frac{e_a\phi}{T_a} J_0(k_\perp
v_\perp/\Omega_a)  \left( \omega - \omega_{*a}^T \right)f_{a0}, 
\label{eq:gk}
\end{equation}
where $\phi$ is the electrostatic potential, $J_0$ is the zeroth order Bessel
function of the first kind, $g_a=f_{a1}+\frac{e_a\phi}{T_a}f_{a0}$ denotes the
non-adiabatic part of the perturbed distribution function,
and the equilibrium distribution function $f_{a0}$ is Maxwellian.
The ratio between the temperature and density gradients is denoted by $\eta_a = d \ln T_a / d \ln n_a$, and we have written $\omega_{*a}^T = \omega_{*a} [1 + \eta_a (x^2 - 3/2)]$, with $x^2 = m_a v^2 / 2 T_a$.
 The system of
equations is closed by the quasi-neutrality condition, 
\begin{equation}
\sum_a \frac{n_a e_a^2}{T_a}\phi = \sum_a e_a \int g_a J_0 \mathrm{d}^3 v.
\label{eq:qn}
\end{equation}
Our argument is based on the energy budget of the instability \cite{Schekochihin2009}.
We define a quantity
\bn P_a = e_a {\rm Im} \left\{ (i v_\| \nabla_\| g_a  -
\omega_{da} g_a ) \phi^\ast J_0 \right\},
\label{Pa}
\en
which is the rate of gyrokinetic energy transfer from the electrostatic field to species $a$. For compactness we have used the notation
$$ \left\{ \cdots \right\} =\int_{-\infty}^{\infty}\frac{\mathrm{d}l}{B} \int (\cdots) d^3v = \int_{-\infty}^{\infty}\mathrm{d}l\sum_{\sigma}\int_0^{\infty}\pi v^3\mathrm{d}v\int_0^{1/B}\frac{\mathrm{d}\lambda}{\left|v_{\|}\right|}(\cdots) $$    
where $\lambda=  v_{\bot}^2/v^2 B$ and $\sigma = v_{\|}/|v_{\|}|$.
We thus multiply the gyrokinetic equation (\ref{eq:gk}) by $e_a J_0 \phi^\ast$, sum
over all species, integrate over velocity space and along the entire field line in ballooning space, $- \infty< l < \infty $, and take
the imaginary part. We note that with a complex mode frequency $\omega=\omega_r + i \gamma $ 
        $$ {\rm Im} \sum_a \left\{ \omega e_a J_0 \phi^\ast g_a \right\} =
        \gamma \sum_a \frac{n_a e_a^2}{T_a} \int\frac{\mathrm{d}l}{B} |\phi|^2 , $$
and
        $$ {\rm Im} \sum_a \left\{ \frac{e_a^2 |\phi|^2}{T_a} J_0^2 (\omega - \omega_{\ast
a}^T) f_{a0} \right\}
        = \gamma \sum_a \frac{n_a e_a^2 }{T_a} \int\frac{\mathrm{d}l}{B}  \Gamma_0(b) 
|\phi|^2 , $$
where $b = k_\perp^2 T_a / m_a \Omega_a^2$ and 
        $$ \Gamma_0(b) = n_a^{-1} \int J_0^2 f_{a0} d^3v < 1. $$
Thus we obtain a relation that describes the energy budget of the fluctuations
        \bn
         -\gamma \sum_a \frac{n_a e_a^2 }{T_a}  \int\frac{\mathrm{d}l}{B} (1- \Gamma_0)
|\phi|^2  = \sum_a P_a, 
         \label{sumPa}
         \en
which is the generalization to an inhomogeneous plasma (in ballooning space) of Eq. (F10) in \cite{Schekochihin2009}. The right-hand side represents the total energy input from the fluctuations into the various species and must be negative for a growing instability.

Now consider a species $a$ with a bounce frequency $\omega_{ba}$ far above the mode frequency, $\omega \ll
\omega_{ba}$, e.g., the electrons in the case of ordinary TEMs or both species in the case of the collisionless trapped-particle instability of Rosenbluth\cite{Rosenbluth1968} and Kadomtsev and Pogutse \cite{Kadomtsev1967}. We further assume
that $0 < \eta_a < 2/3$ (so that $\omega_{*a}$ and $\omega_{*a}^T$ have the same sign for all energies) and that $\overline{\omega}_{da}$ has the same sign for all orbits. Thus ordering $\omega \sim \omega_{*a} \ll k_{\|}\left(T_a/m_a\right)^{1/2}$ we can expand the distribution
function, $g_a = g_{a0} + g_{a1} + \cdots$ and obtain
        $$ g_{a0} = \frac{e_a \overline{J_0 \phi}}{T_a} \frac{\omega - \omega_{\ast
a}^T}{\omega - \overline{\omega}_{da}} f_{a0}, $$ and 
        $$ i v_\| \nabla_\| g_{a1} = (\omega - \omega_{\ast a}^T ) \frac{e_a}{T_a} 
        \left( J_0 \phi - \frac{\omega - \omega_{da}}{\omega - \overline{\omega}_{da}}
\overline{J_0 \phi} \right) f_{a0}.$$
Here we have neglected the passing particles, whose response is a factor
$\omega/k_\|v_{Ta}\ll 1$ smaller than that of the trapped ones.
Substituting these results in the expression (\ref{Pa}) for the energy transfer gives
        \begin{equation}
         P_a = \frac{e_a^2}{T_a} {\rm Im} \left\{ (\omega - \omega_{\ast a}^T )
        \left( \overline{|J_0 \phi |^2} - \frac{\omega |\overline{J_0 \phi}|^2}{\omega -
\overline{\omega}_{da}} \right) f_{a0} \right\}.
\label{eq:Pa gammafinite}
\end{equation}
Finally, we consider the limit of this expression when marginal stability is
approached, $\gamma \rightarrow 0+$, where we obtain
        \begin{equation}
         P_a = \frac{\pi e_a^2}{T_a} \left\{ \delta(\omega - \overline{\omega}_{da})
\overline{\omega}_{da} 
        (\overline{\omega}_{da} - \omega_{\ast a}^T) 
        |\overline{J_0 \phi}|^2 f_{a0} \right\}.
        \label{eq:Pa gamma0} 
        \end{equation}
If $\omega_{\ast a}^T$ and $\overline{\omega}_{da}$ are of opposite signs, $P_a >
0$ and energy flows from the electric field fluctuations to plasma species $a$,
which therefore exerts a stabilizing influence. Consequently, for instabilities with such low frequencies that $\omega \ll
\omega_{ba}$ for {\em all} species,
we find that $ \sum_a P_a > 0$, which is in contradiction to Eq. (\ref{sumPa}) at the
point of marginal stability, implying the non-existence of a marginal stability point and consequently the absence of an instability.
The case where the real part of the frequency vanishes, $\omega_r=0$, requires slightly more care, since the resonance then occurs at zero energy and consequently all $P_a=0$, so that an instability cannot be ruled out by Eq. (\ref{eq:Pa gamma0}). However, from Eq. (\ref{eq:Pa gammafinite}), we obtain for $\omega_r=0$
$$P_a = \frac{e_a^2}{T_a} \left\{ \left[ \gamma \overline{|J_0 \phi|}^2  + \frac{\gamma}{\overline{\omega}_{da}^2 +\gamma^2}\left( -\omega_{\ast a}^T \overline{\omega}_{da} -\gamma^2\right) \overline{|J_0 \phi |^2} \right] f_{a0} \right\},$$
where for small but finite $\gamma$ all terms are positive, again in contradiction to Eq. (\ref{sumPa}); therefore a mode with $\omega_r=0$ at marginal stability cannot exist.
Hence we conclude that the collisionless trapped-particle mode is absent -- i.e., there is no
instability with frequency far below the ion bounce frequency. This conclusion is an
extension of the result by Rosenbluth to an arbitrary number of particle species,
finite $k_\perp \rho_a$, finite temperature gradients up to $\eta_a < 2/3$, and finite values of $\omega/\omega_{da}$.

If only the electrons have a bounce frequency that exceeds $\omega$ but $\omega \sim \omega_{bi}$, then we cannot
rule out instability by this argument, but we can say something about the nature of a possibly occuring mode. We proceed from the gyrokinetic equation (\ref{eq:gk}) and treat it as we did to obtain Eq. (\ref{sumPa}), only that we do not sum over the species. We then find at the point of marginal stability
$$
P_a=-\omega_r {\rm Im}\left\{e_a g_a J_0 \phi^\ast \right\}\equiv -\omega_r Q_a.
$$
The distribution function $g_a$ appearing in this quadratic form $Q_a$ can be obtained from the solution of the gyrokinetic equation (\ref{eq:gk}) given in Refs.~\cite{Connor1980} and \cite{Tang1980} (correcting for misprints). In the region of velocity space corresponding to trapped particles, $\lambda> 1/B_{\rm max}$, where $B_{\rm max}$ denotes the maximum field strength along the field line, the solution is 
\begin{equation}
\sum_{\sigma}g_{a,t}(l)=\frac{2 e_a f_{a0}}{T_a}\frac{\omega -\omega^T_{*a}}{\sin \left(M(\omega,l_1, l_2)\right)}\int_{l_1}^{l_2}\frac{\mathrm{d}l'}{\left|v_{\|}\right|}\phi J_0 \cos\left(M(\omega,l_1, l_l)\right)\cos\left(M(\omega,l_u, l_2)\right),
\end{equation}
where $l_1(\lambda)$ and $l_2(\lambda)$ are the bounce points (defined by $\lambda B = 1$) immediately surrounding $l$, and where we have written $l_u=\max(l, l')$ and $l_l=\min(l, l')$ and defined 
\begin{equation*}
M(\omega,a,b)=\int_{a}^{b}\frac{\mathrm{d}l'}{\left|v_{\|}\right|}\left(\omega-\omega_{da}\right).
\end{equation*}
If the growth rate $\gamma$ is taken to be positive the solution in the untrapped region is given by
\begin{equation}
\sum_{\sigma}g_{a,p}(l)=\frac{e_a f_{a0}}{\pi T_a}(\omega -\omega^T_{*a})\int_{-\infty}^{\infty}\frac{\mathrm{d}t}{\omega - t}\int_{-\infty}^{\infty}\frac{\mathrm{d}l'}{\left| v_{\|}\right|}\phi J_0 \cos\left(M(t,l, l')\right).
\label{passing solution}
\end{equation}

The quadratic form $Q_a$ can be written as a sum of the contributions from trapped and passing particles of each species separately, $Q_a = Q_{at} + Q_{ap}$. Substituting the solution (\ref{passing solution}) for passing particles gives
\begin{equation}
Q_{ap}(\omega)=\mathrm{Im} \; \frac{e_a^2 n_a}{T_a v_{Ta}\pi^{3/2}}\int_{-\infty}^{\infty}\frac{\mathrm{d}t}{\omega - t}\int_0^{\infty}\mathrm{d}x(t -\omega^T_{*a}) e^{-x^2}x \int_{0}^{1/B_{max}}\mathrm{d}\lambda \sum_{j=\cos, \sin}\psi^*_j(x, \lambda, t)\psi_j(x, \lambda, t),
\end{equation}
where $v_{Ta} = (2 T_a/m_a)^{1/2}$ is the thermal velocity and
%\begin{align*}
%\psi_{\cos}(\lambda, t)&=\int_{-\infty}^{\infty}\frac{\mathrm{d}l J_0 \phi}{B \sqrt{1-\lambda B}} \cos\left(M(t,0,l), \right) \\
%\psi_{\sin}(\lambda,t)&=\int_{-\infty}^{\infty}\frac{\mathrm{d}l J_0 \phi}{B \sqrt{1-\lambda B}} \sin\left( M(t,0,l). \right)
%\end{align*}
$$ {\psi_{\cos}(x, \lambda, t) \choose \psi_{\sin}(x, \lambda,t)}=\int_{-\infty}^{\infty}\frac{\mathrm{d}l J_0 \phi}{\sqrt{1-\lambda B}} {\cos \choose \sin} \left(M(t,0,l) \right).$$
Therefore, at marginal stability, where $\omega$ has an infinitesimal positive imaginary part, we have
\begin{equation}
Q_{ap}(\omega)=-\frac{e_a^2 n_a}{T_a v_{Ta}\sqrt{\pi}}\int_0^{\infty}\mathrm{d}x(\omega -\omega^T_{*a}) e^{-x^2}x
\int_{0}^{1/B_{max}}\mathrm{d}\lambda \sum_{j=\cos, \sin} \left| \psi_j(x, \lambda,\omega)  \right|^2 .
\label{Qap}
\end{equation}
In ballooning space, there is an infinity of trapping regions along the field line, which are periodic in a tokamak but irregularly distributed in a stellarator. When calculating the contribution from the trapped particles to the quadratic form $Q_a$ we need to sum over all these trapping wells, and then obtain
\begin{equation*}
Q_{at}(\omega)=\mathrm{Im} \; \frac{2 e_a^2 n_a}{T_a v_{Ta}\sqrt{\pi}}\int_0^{\infty} \mathrm{d}x(\omega -\omega^T_{*a}) e^{-x^2}x \int_{1/B_{min}}^{1/B_{max}}\mathrm{d}\lambda 
\end{equation*}
\begin{equation} \times
\sum_{\mathrm{wells}}\frac{1}{\sin(M(\omega,l_1, l_2))} \int_{l_1}^{l_2}\frac{\mathrm{d}l \phi^* J_0}{\sqrt{1-\lambda B}} \int_{l_1}^{l_2}\frac{\mathrm{d}l' \phi' J'_0}{\sqrt{1-\lambda B'}}\cos(M(\omega,l_1, l_l))\cos(M(\omega,l_u, l_2)).
\end{equation}
Near marginal stability, an imaginary contribution arises due to the zeros of the sine in the denominator, and when splitting the cosines symmetrically we find
\begin{equation}
Q_{at}(\omega)=-\frac{2 \sqrt{\pi}e_a^2 n_a}{T_a v_{Ta}}\sum_{m=-\infty}^{\infty}
\int_0^{\infty}\mathrm{d}x (\omega -\omega^T_{*a}) e^{-x^2}x
\int_{1/B_{min}}^{1/B_{max}}\mathrm{d}\lambda \sum_{\mathrm{wells}}
\delta \left( M(\omega,l_1, l_2)-m\pi \right)\left|\psi_t(x, \lambda,\omega) \right| ^2,
\label{Qat}
\end{equation}
with
\begin{equation*}
\psi_t(x, \lambda,\omega)
= \int_{l_1}^{l_2}\frac{\mathrm{d}l \phi J_0}{\sqrt{1-\lambda B}}\cos(M(\omega,l_1, l)).
\end{equation*}
We now note that for all species $a$ the forms $Q_{at}$ and $Q_{ap}$ have the character of a weighted average over $x$ of $(\omega -\omega^T_{*a})$, due to the positive-definiteness of the other factors. Thus we can write
	\begin{equation}
	 P_a = \omega \int_0^\infty (\omega -\omega^T_{*a}) \mathrm{Pos}_a(x,\omega) dx, 
	\label{weighted average}
	\end{equation}
where $\mathrm{Pos}_a(x,\omega)$ is a positive definite function. 
If we now assume the mode travels in the electron diamagnetic direction, i.e. $\omega\omega_{*e}>0$, we know from Eq. (\ref{eq:Pa gamma0}) that $P_e=0$ due to the lack of resonance.  Consequently Eq. (\ref{sumPa}) then implies that $P_i=0$ at the point of marginal stability. However, from Eq. (\ref{weighted average}) we obtain $P_i>0$ since $\omega\omega_{*i}<0$, which again implies the non-existence of the marginal stability point and the absence of this particular mode.
Thus any unstable mode that could arise with $\omega \sim \omega_{bi}$ must propagate in the ion direction at marginal stability and as a consequence from Eq. (\ref{eq:Pa gamma0}) draw energy from the ions rather than the electrons ($P_e>0$ follows from Eq.(\ref{eq:Pa gamma0}), with Eq. (\ref{sumPa}) then implying $P_i<0$).
There are thus no ordinary TEMs, which tend to cause much of the
transport observed in tokamaks. We also note that in the usual treatment of the ``ubiquitous'' mode of Coppi and Rewoldt \cite{Coppi1974a} $\omega_{*a}\cdot \overline{\omega_{da}}<0$ implies stability as well.
These conclusions hold as long as $0<\eta_a <2/3$ for all species, and collisions may be ignored, but the dissipative TEM could still be unstable.

Since this argument is essentially only based on the requirement of quasineutrality
and an analysis of the energy budget, it is independent of all geometric details of
the magnetic field except the condition that the bounce-averaged curvature should be
favorable,  $\partial J / \partial \psi < 0$, for all orbits. This requirement can
also be satisfied in other omnigenous configurations \cite{Cary1997,Hall1975}. In a
tokamak, for example, it is achieved if the pressure gradient is steep enough to
cause drift reversal of all trapped particles \cite{Connor1983}, but in practice
such a steep pressure gradient necessitates taking account of electromagnetic
effects. However, if MHD ballooning modes are stabilized by negative magnetic shear
(according to the tokamak definition), it is conceivable that the stabilization of
trapped-electron modes could help explain the transport reduction observed in
internal transport barriers.

It is, of course, difficult to achieve exact quasi-isodynamicity, but one expects
that the drive for trapped-particle modes should become weak if {\em most} orbits
satisfy $\omega_{*e}\cdot \overline{\omega_{de}}<0$.
One would expect that even approximately quasi-isodynamic stellarators should have
relatively small trapped-particle instability growth rates, particularly if central
fueling is accessible through pellet injection so that a stabilizing density
gradient can be achieved. Finally, it should be mentioned that inverting the density
gradient in a tokamak, so as to reverse the sign of $\omega_{*e}$, has long been
known to make the collisionless trapped-electron mode less unstable
\cite{Tang1975,Horton1976}, because there are then fewer electrons with
$\omega_{*e}\cdot \overline{\omega_{de}}>0$. However, this stabilization is
incomplete since in a typical tokamak there are always electrons with both signs of
$\overline{\omega_{de}}$. 

In summary, whereas in tokamaks most of the transport in the core tends to be driven
by ITG and ETG modes, and by modes driven unstable by trapped electrons, the latter are stable in
quasi-isodynamic stellarators in the electrostatic and collisionless limit, if $0 < \eta_a < 2/3$, and so are also all such instabilities with frequencies below the ion bounce frequency.

\begin{acknowledgments}
One of the authors (J.W.C.) gratefully acknowledges funding from IPP Greifswald.
\end{acknowledgments}

%\bibliography{library}{}
%\bibliographystyle{plain}

\end{document}